\documentclass[twocolumn,showpacs,amsmath,amssymb,amsfonts,superscriptaddress]{revtex4}

\PassOptionsToPackage{caption=false}{subfig}
\usepackage{subfig}
\usepackage{enumerate}
\usepackage{graphicx}
\usepackage{color}
\usepackage{dcolumn}
\usepackage{bm}

\begin{document}

\author{Shahin Mobarakabadi}
\author{Ehsan Nedaaee Oskoee}
    \email{nedaaee@iasbs.ac.ir}
    \affiliation{Institute for Advanced Studies in Basic Sciences, Gava Zang, Zanjan 45195-159, Iran}
\author{Matthias Schr\"oter}
    \affiliation{Max Planck Institute for Dynamics and Self-Organization (MPIDS), 37077 G\"{o}ttingen, Germany}
\author {Mehdi Habibi}
    \affiliation{Institute for Advanced Studies in Basic Sciences, Gava Zang, Zanjan 45195-159, Iran}

\title{Granular transport in a horizontally vibrated sawtooth channel}

\date{\today}

\begin{abstract}
We present a new mode of transport of spherical particles in a horizontally vibrated channel with
sawtooth shaped side walls.  The underlying driving mechanism  is based on an interplay of
directional energy injection transformed by the sidewall collisions and density dependent 
interparticle collisions. Experiments and matching numerics show that the average particle velocity
reaches a maximum at  60 \% of the maximal filling density. Introducing a spatial phase shift between 
the channel boundaries increases the transport velocity by an order of magnitude.
\end{abstract}



\maketitle

\section{Introduction}
Brownian motors extract useful work from a noisy environment
by means of a broken spatial symmetry \cite{haenggi:09,reimann:02}. The concept of Brownian motors 
is of great importance in cell biology and nanotechnology, however their underlying principle is not limited to 
thermal noise and can therefore be also implemented in macroscopic setups.  One example for such an 
athermal, macroscopic, and  noisy environment are granular gases \cite{brilliantov:04}. 

Due to the dissipative collisions between particles, granular gases require constant external driving 
which is in most cases provided by shaking the container. Work can then e.g.~be extracted by means of a rotational ratchet, 
where the symmetry is broken by different coatings on the two sides of each vane \cite{eshuis:10}. Or probe particles
with an asymmetric shape can be set into translational \cite{cleuren:07,costantini:07} or rotational  \cite{balzan:11} motion.

Another class of granular Brownian motors converts the random motion of the particles into a directed flow. 
This can either be implemented by breaking the symmetry of the driving  \cite{gallas:92,grochowski:04,blaschke:13}, or by 
breaking the spatial
symmetry of the container boundaries \cite{derenyi:98,farkas:99,levanon:01,farkas:02,wambaugh:02,bae:04,heckel:12,chen:13}.
All setups in the latter group include a sawtooth shaped base plate which is shaken vertically to drive the granular gas. Their
phenomenology includes height-dependent flow directions \cite{derenyi:98,farkas:99,levanon:01,bae:04}, 
segregation of binary mixtures  \cite{farkas:02,wambaugh:02}, and rotational motion of the circular base \cite{heckel:12}.

In this paper we decouple the direction of driving from gravity:
our system is a horizontally vibrated channel where 
sawtooth shaped side walls break the symmetry. 
This geometry  allows us to control the average particle density, a feature which can not be realized with
vertically shaken cells with sawtooth floor. 
We observe unidirectional transport with a non-monotone 
density dependence. In section 2 we discuss the experimental setup and results. 
In section 3 we introduce and validate our numerical simulations
which are then used in section 4 to gain insight into the transport mechanisms. 

\section{Experimental results}
\begin{figure}[t]
\subfloat[]{\label{setup:Exp}\includegraphics[trim = 0mm 260mm 360mm 0mm, clip=true, scale=0.14, angle=90]{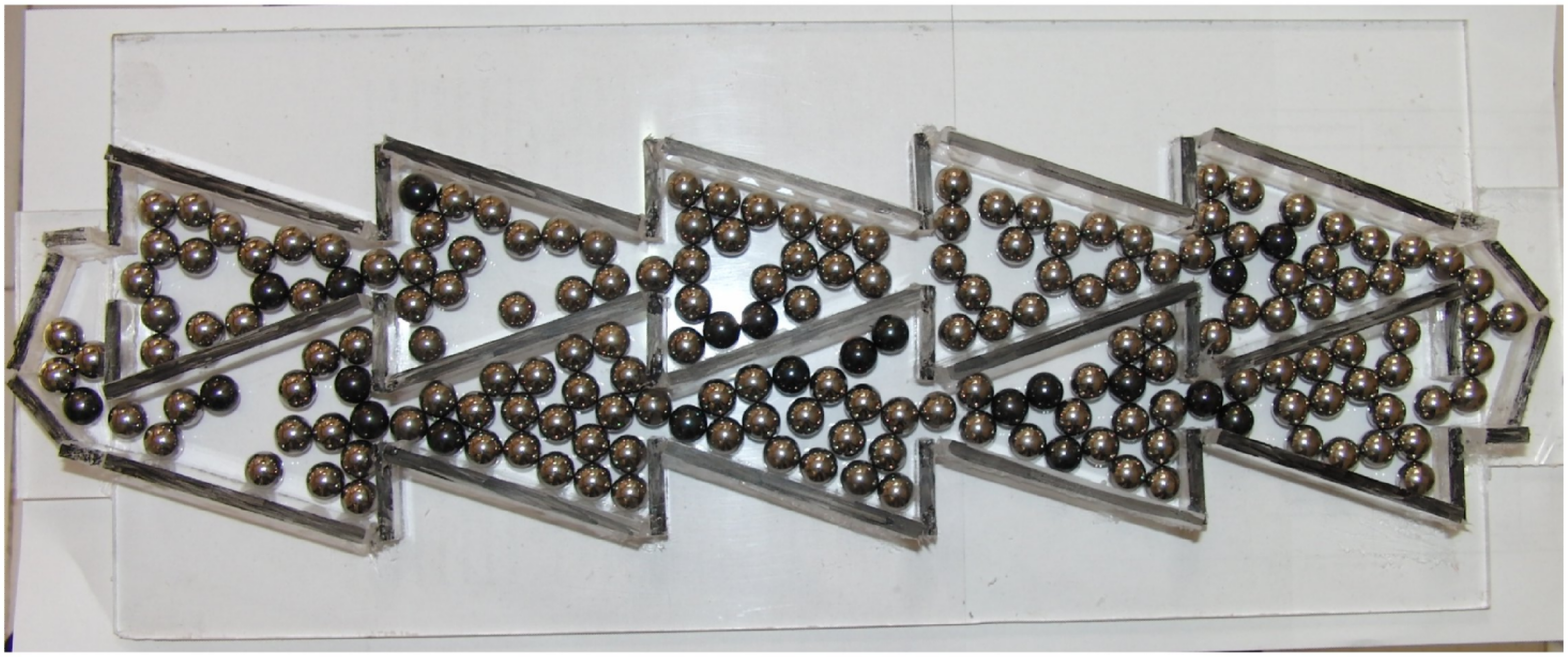}}
\subfloat[]{\label{setup:dim}\includegraphics[trim = 80mm 0mm 0mm 0mm, clip=true, scale=0.20,angle=0]{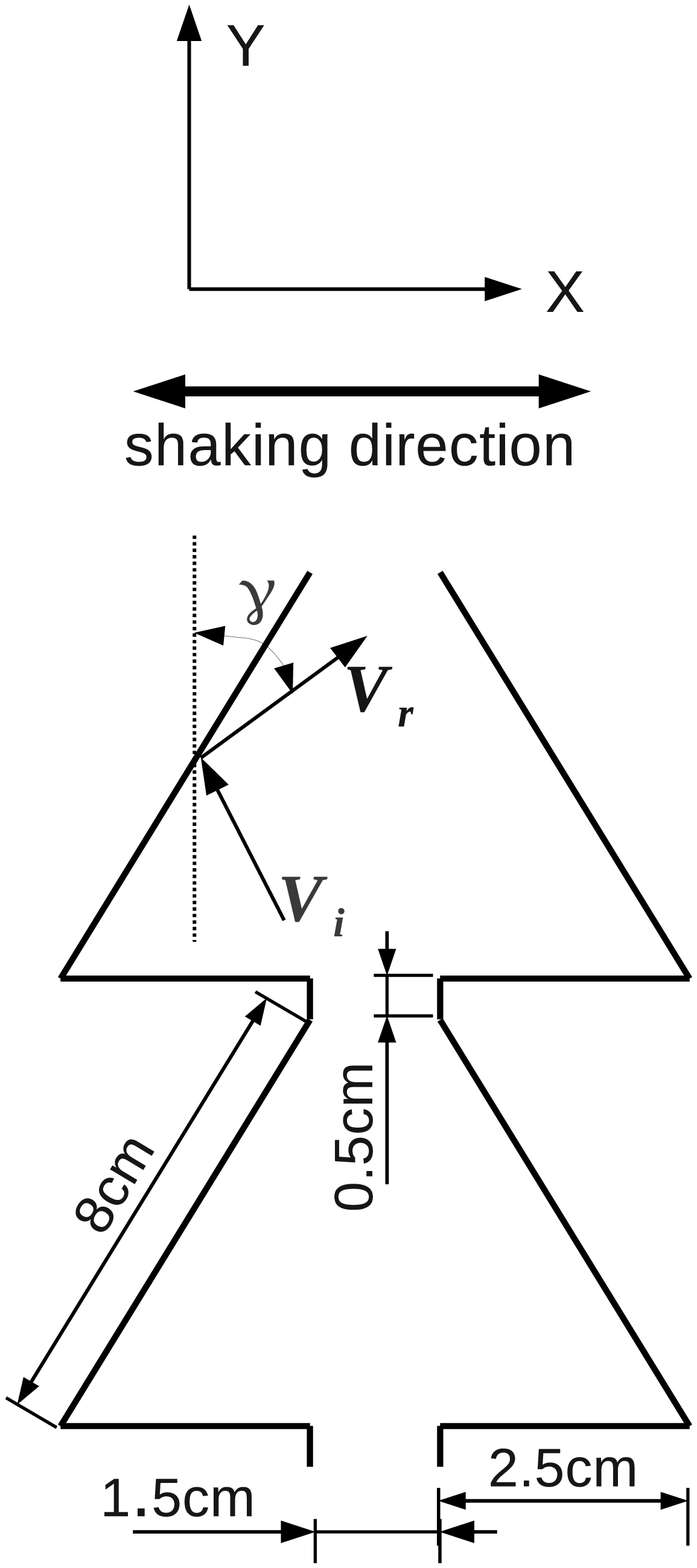}}
\subfloat[]{\label{setup:simu}\reflectbox{\includegraphics[trim = 0mm 0mm 0mm 10mm, clip=true, scale=0.31, angle=90]{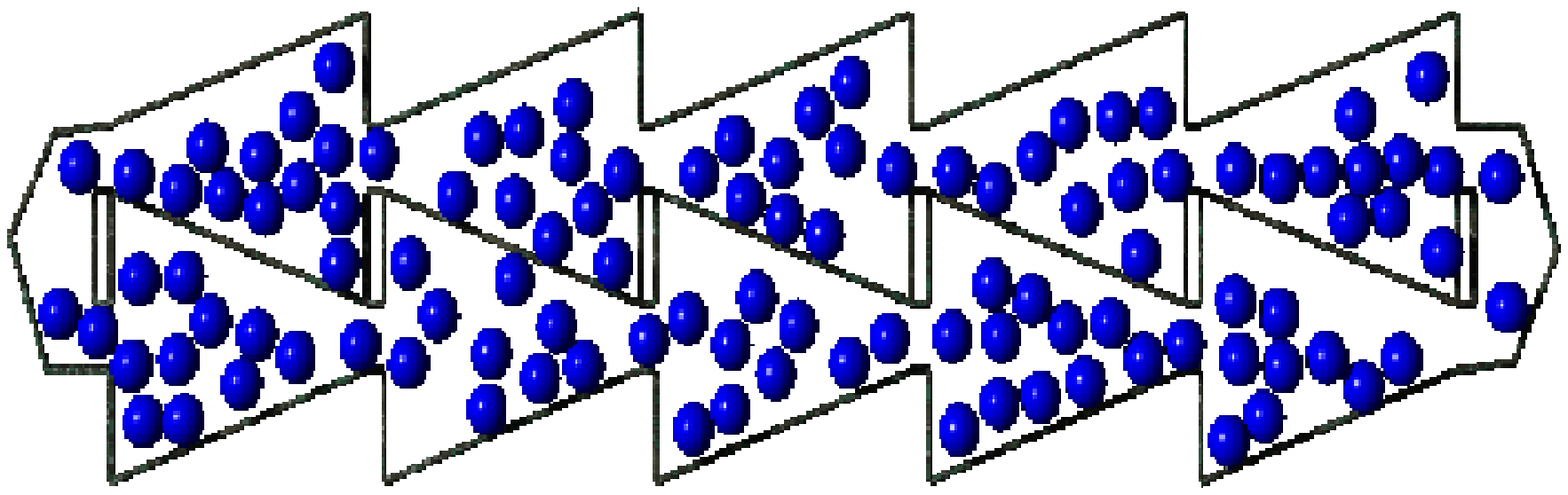}}}
\caption{(Color online)
Sawtooth channel.   Top view (a),  and  dimensions  (b) of the experimental setup 
and visualization of the setup used in the simulations (c).
Particles are transported in the negative Y-direction in panel (b). 
}
\label{fig:setup}
\end{figure}

The experimental setup consists of two sawtooth corridors which are connected by two small pentagonal
cells at each end of the corridors, as shown in Fig.~\ref{setup:Exp}.
Each corridor consists of five triangular cells; both floor and walls are made from plexiglass.
The channels are mounted horizontally and shaken in X direction (c.f.~fig.~\ref{setup:dim})
using an off center pulley which is driven by an AC motor.
The oscillations amplitude  is  $7.5 \pm 0.5$~mm, the driving frequency 5 Hz.

The channel is filled with steel spheres of radius 6 mm. The maximum number of spheres per triangular cell
is 21, this value defines the maximal area fraction  $\phi_{c}$. 
Ten percent of the beads have been colored in order to use them as tracers, 
this has been achieved by oxidizing them in a fire for one minute. 
We checked with a simple bouncing experiment that their restitution coefficient
was within the error bar of untreated steel spheres. 
The average speed of the beads is then determined by measuring the rate at which tracers enter a given cell; 
the sample time is chosen so that the average tracer has completed at least two full cycles.

\begin{figure}[t]
\begin{centering}
\includegraphics[width=8.8cm]{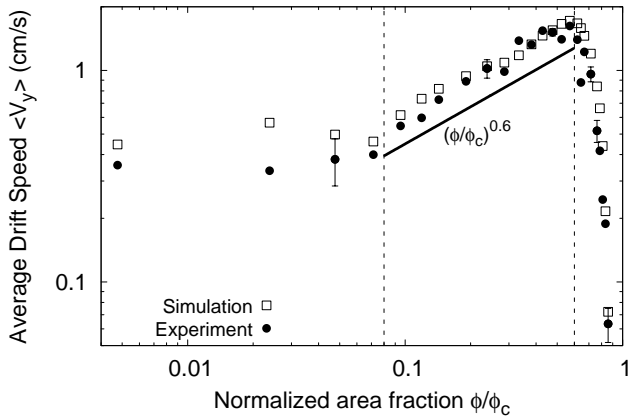}
\end{centering}
\caption{Average drift speed as a function of the rescaled area fraction, measured both in experiment  and simulation.
Vertical dashed lines indicate the transition between the dilute (left), cooperative (middle) and jamming 
regime (right).
The three example error bars are standard deviations over 10 cycles. 
The simulation time corresponds to the average particle performing four cycles.
In both cases the oscillations amplitude is  $7.5$ mm and the driving frequency 5 Hz.
}
\label{fig:logspeed}
\end{figure}

Figure \ref{fig:logspeed} shows the average drift speed as a function of the renormalized area fraction
($\phi/\phi_{c}$) where $\phi$ is the number of spheres per cell.
Three different  regimes can be distinguished. For low area fractions
($0.005 \leq \phi/\phi_{c} < 0.08 $) the particles move in negative Y direction
(i.e.~downward in fig.~\ref{setup:dim})
with a velocity that is independent of $\phi$. As here are
on average less than 2 beads in a cell, particle-particle collisions are rare.
We refer to this range as the {\it dilute regime}.

For intermediate area fractions ($0.08 \leq \phi/\phi_{c} < 0.6 $)
the average speed increase monotonously. Fitting a  line to
the experimental data in figure \ref{fig:logspeed}
shows that $\langle V_y \rangle \sim (\phi/\phi_c)^{0.6 (\pm 0.05)}$
in the intermediate regime. This effect delineates the {\it cooperative regime}.

For even higher densities ($\phi/\phi_{c} \geq 0.6 $)
the drift speed decreases sharply.
As the system approaches the fluid-solid transition density,
we refer to this as the {\it jamming regime}.

\section{Simulation method}
To elucidate the transport mechanism we have performed
molecular dynamics simulations. 
The normal force between two grains $\textit{i}$ and $\textit{j}$,
$\mathbf{F}_{n_{i,j}}$, is given by
\begin{equation}
\textbf{F}_{n_{i,j}}=f(\frac{\delta_{i,j}}{d})(k_n\delta_{i,j}\textbf{n}_{i,j}-
\gamma_{n}m_{eff}\textbf{v}_{n_{i,j}}).
\label{eq6}
\end{equation}
Here $\textbf{v}_{n_{i,j}}$is  the normal component of the relative velocity between the grains
(evaluated at the contact point). $\textbf{n}_{i,j}$ is the unit vector
corresponding to the distance ${r}_{\textit{i,j}}$ between the two grain centers.
$d$ is the particle diameter, $m_{eff}$  the effective mass $m_{i}m_{j}/(m_{i}+m_{j})$, and
 $\delta_{i,j}={d}-{r}_{\textit{i,j}}$ is the normal compression at the contact.
$f(x)$ equals $\sqrt{x}$ which models Hertzian contacts.
The density of the grains is set to 7.87 g/cm$^3$.
The numerical values of the
elastic constant $k_n$ and the viscoelastic constant $\gamma_{n}$
are listed in table \ref{TAB1}; they have been determined by numerically matching
the experimental results of a steel sphere bouncing of a steel plate (respectively a plexiglass 
plate for the particle wall collisions described below).

We do neglect the the viscoelastic tangential interaction between grains, 
this assures that particles do not experience any vertical forces and remain therefore in 
contact with the bottom surface of the channel. The agreement shown below between our 
simulations and experiments justifies this decision by hindsight.  

\begin{table}[t]
\begin{center}
\begin{tabular}{c|c||c|c}
Coefficient & Numeric Value & Coefficient & Numeric Value\\
\hline
 $k_{n}$ & $10^7$ $\rm g/s^2$ & $k_{nw}$ & $10^7$  $\rm g/s^2$\\
\hline
 $\gamma_{n}$ & $5 \times 10^3$  1/s & $\gamma_{nw}$ & $1.3 \times 10^4$ 1/s\\
\hline
$\mu_s$ & $0.5$ & $\mu_{d}$ & $0.4$\\

\end{tabular}
\end{center}
\caption{The numerical values of the contact model coefficients.}
\label{TAB1}
\end{table}

The same equations are used  to model the interaction between grains and the channel walls, the
coefficients have an added subscript w. Viscoelastic tangential forces between the grains and
side walls are neglected too. 
The mass of the container is set to be infinity, consequentially $m_{eff}=m$.

For the evaluation of the tangential force $\mathbf{f}$ exerted by the channel bottom on a particle we follow 
a method proposed by  Kondic \cite{kondic:99} which distinguishes between sliding and rolling contacts.
In a first step the contact is assumed to be rolling with $\mathbf{f}=m\mathbf{a}$. 
Taking the rotation into account, the acceleration $\mathbf{a}$ of the sphere 
can be computed as $\mathbf{a}=2/7 \mathbf{a}_s$ with $\mathbf{a}_s$ being the acceleration of the surface.
If the no-sliding condition $|\mathbf{f}| \leq f_{max} = \mu_s m g$
($mu_s$ is the static coefficient of friction)  is satisfied, this is the final result. 

Otherwise, the sliding contact has to be evaluated using the dynamic friction coefficient  $\mu_{d}$:
\begin{equation}
 \mathbf{f} = -\mu_{d} mg  \dfrac{\mathbf{v}_w}{|\mathbf{v}_w|}
\end{equation}
where $\mathbf{v}_w$ is the velocity of the contact point relative to the substrate.
The values of $\mu$ used are listed in table \ref{TAB1}, they are only justified by their
ability to reproduce the experimental results.

The shaking of the container is modeled by a sinusoidal excitation
with an oscillation frequency $\omega$  and  an amplitude ${x}_{0}$.
As the simulations are performed in the referential frame of the container, 
a force $- \omega^2 {x}_{0} \cos{(\omega t)}$ is added to each particles center of mass.
The code is written in C++
using a 5th order predictor-corrector algorithm for numerical
integration of the equations of motion.
The time step increment ($\Delta t$) is set to $10^{-5}$ s in all simulations;
the simulation time is chosen so that the average particle has performed four full cycles which requires at maximum $3.6 \times 10^7$ time steps.

Figure \ref{fig:logspeed} shows that  the results of
the simulation are in good agreement with the experiment.
Therefore we will use our simulations to gain insight into the driving mechanism in the three regimes.

\section{Transport Mechanism}
\begin{figure}[t]
\subfloat[]{\label{setup:oblique}\includegraphics[scale=0.28, angle=90]{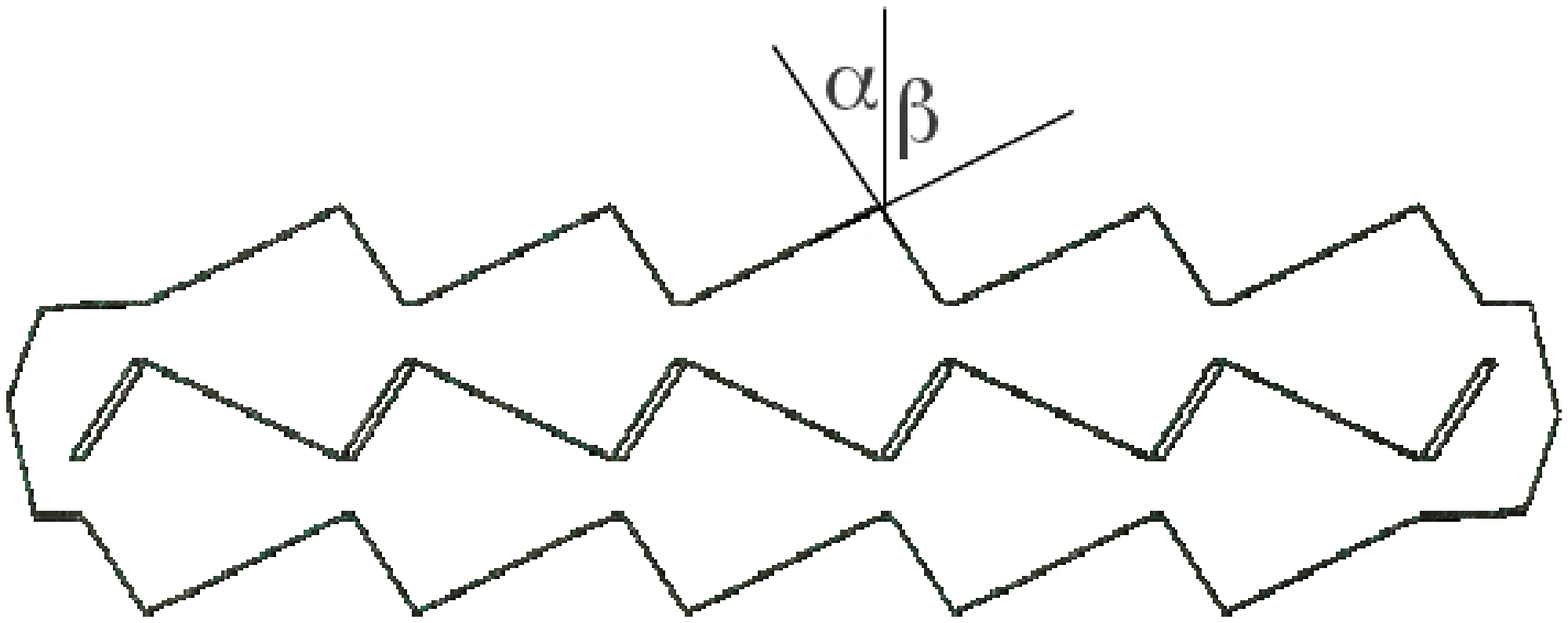}}
\subfloat[]{\label{setup:shift}\includegraphics[trim = 20mm 40mm 40mm 0mm,scale=0.28]{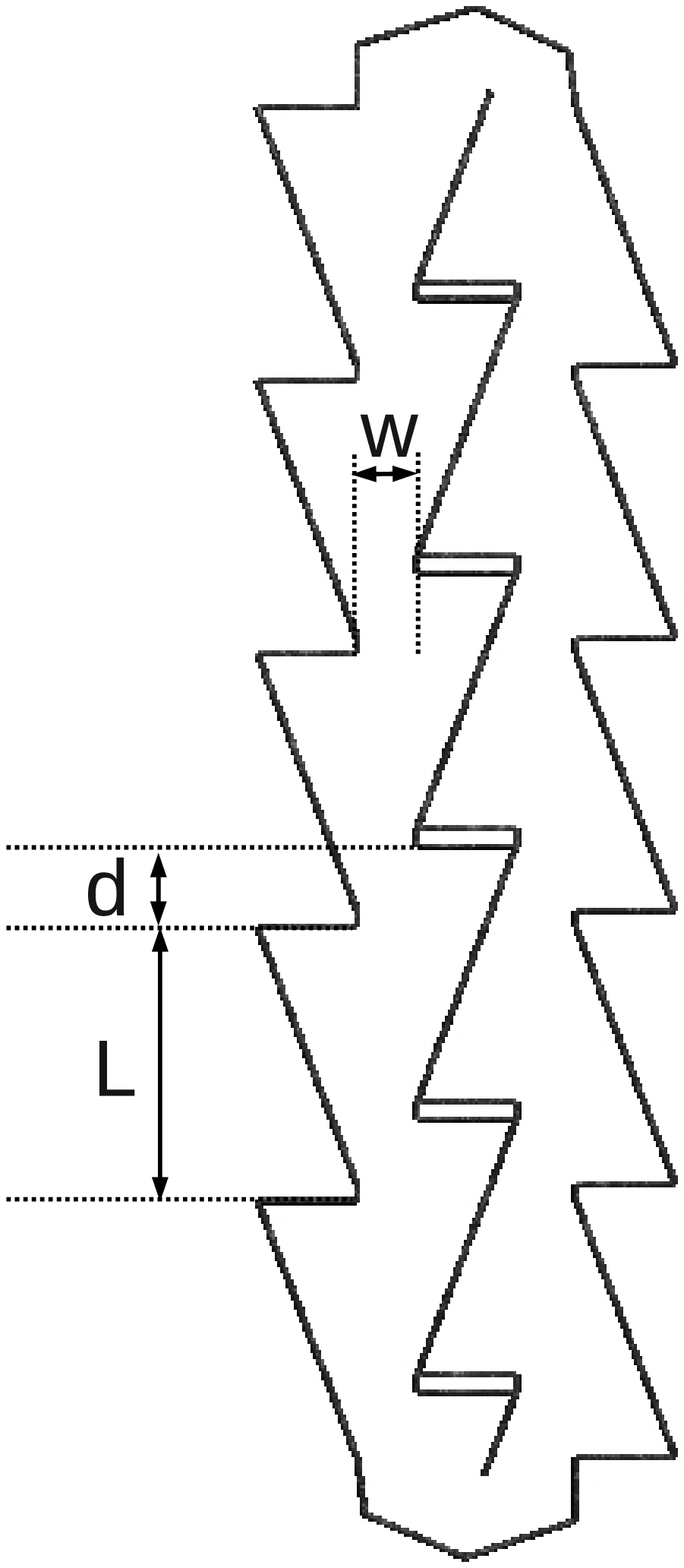}}
\caption{Modified boundary geometries.
Channel a) interpolates between a sawtooth and a triangular shape by increasing the angle $\alpha$.  
$\beta$ is kept constant at 72 degree as in figure \ref{fig:setup}.
Channel b) introduces a relative displacement $d$ of the two sidewalls. 
}
\label{fig:changed_geometry}
\end{figure}

Understanding the transport mechanism depicted in figure \ref{fig:logspeed} requires answers to the following three questions: 
\begin{enumerate}[(A)]
\item Why are the particles moving unidirectionally at all?
This is best discussed in the dilute regime.
\item Why does the drift velocity increase in the cooperative regime?
\item Why does the drift velocity go down at even higher densities?
\end{enumerate}

\subsection{Particles moving in the dilute regime}
The dilute regime is specifically apt to study the underlying transport mechanism as it allows us to ignore inter-particle 
collisions. First, it should be pointed out that dissipative collisions are not a necessary condition to have a 
finite drift velocity; the mechanism is therefore different from fluxes seen in granular gases going through beveled pores  
\cite{shaw:07}. This has been tested by short simulations where $\gamma_n$ and $\gamma_{nw}$ have been set to zero.

\begin{figure}[t]
\begin{centering}
\includegraphics[width=8.8cm]{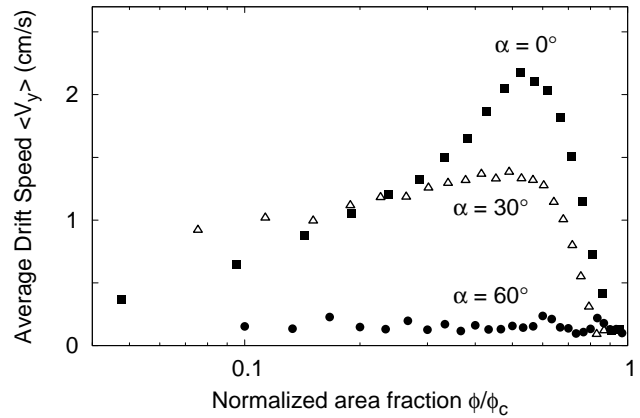}
\end{centering}
\caption{The average drift speed depends on the asymmetry of the channel.
These numerical data are measured in setup \ref{setup:oblique} at a shaking frequency of 6.4 Hz.
and an the oscillations amplitude of  $7.5$ mm.
 }
\label{fig:sidewall_angle}
\end{figure}

Secondly, like in other ratchet systems the symmetry breaking of the  boundary is a necessary condition 
to obtain a finite $\langle v_y \rangle $. If we increase in our simulations the angle $\alpha$ shown
in Figure \ref{setup:oblique}, we change the channel profile from a sawtooth shape to a triangular one.
This goes together with a reduction of the average drift velocity towards zero as shown in 
figure \ref{fig:sidewall_angle}.

\begin{figure}[ht]
\begin{centering}
\includegraphics[width=8.8cm]{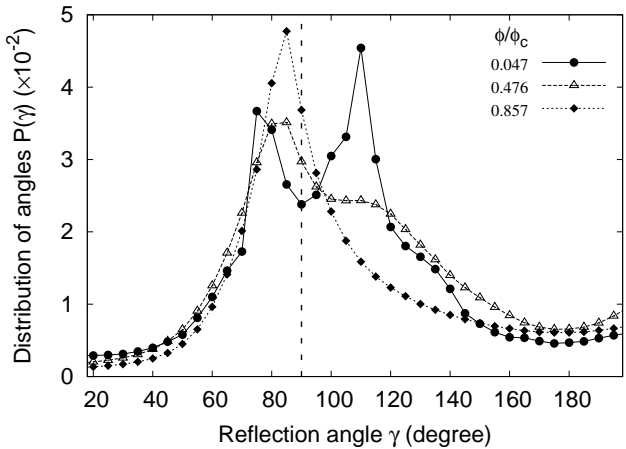}
\end{centering}
\caption{Angular distribution of particle velocities after collisions with the sawtooth side walls. 
The angle $\gamma$ is
defined in figure \ref{setup:dim}. Values of $\gamma$ between 90 and 198 degrees make the particles go 
into the direction of the average drift.
Numerical data measured with an oscillations amplitude of  7.5 mm and a driving frequency of 6.4 Hz.
}
\label{fig:dist-alpha}
\end{figure}

We therefore conjecture the following driving mechanism: spheres gain kinetic energy in x-direction
due to the frictional contacts with the horizontally vibrating base. This kinetic energy is then 
converted by the sidewall collisions into a directed motion in negative  y-direction. To substantiate this 
conjecture, we have computed the distribution $P$ of particle direction with respect to the Y-axis $\gamma$,
measured after the particles collided with the sidewall  (c.f.~Fig.~\ref{setup:dim}).  
As it can be seen in figure \ref{fig:dist-alpha}, the majority of post collisional velocities point in negative 
y-direction which corresponds to $\gamma$ values larger than 90 degree. For $\phi/\phi_c$ = 0.047 more than 60 \% of
post collision vectors fall into this range. Our system seems therefore in the dilute limit amenable for the 
analysis presented in \cite{bae:04}, which would explain why for our comparatively elastic steel particles
and large ratchet angles no reversal of the transport direction was observed.

\subsection{Cooperative regime}
For values of  $\phi/\phi_c > 0.08$, the average drift velocity increases proportional to  $(\phi/\phi_c)^{0.6}$ 
as shown in figure \ref{fig:logspeed}. However, the post sidewall collision angle distribution $P(\gamma)$  even becomes 
more balanced, at $\phi/\phi_c$ = 0.476 only 58 \%  of the vectors point in negative y-direction. 
Therefore the mechanism behind the increase of  $\langle v_y \rangle $ has to be something different. We hypothesize
that the increased density keeps the particles longer in close proximity to the narrow passage between cells. 

\begin{figure}[t]
\begin{centering}
\includegraphics[width=8.8cm]{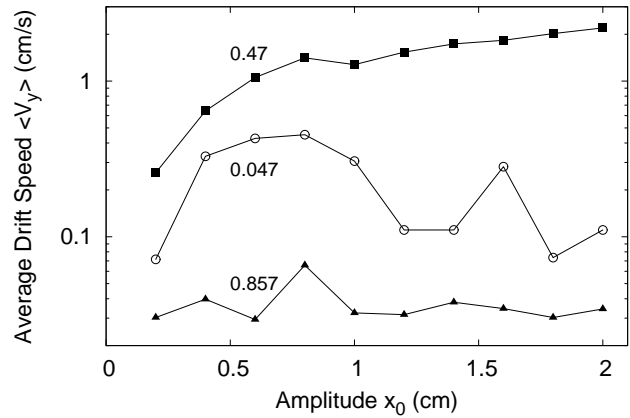}
\end{centering}
\caption{ Average drift speed versus shaking amplitude for three different values of $ \phi/\phi_{c}$.
The drift amplitude increases only in the cooperative regime with the driving strength. 
These numerical results are measured in the standard setup displayed in figure \ref{setup:simu}
with a driving frequency of 5 Hz.
}
\label{fig:amplitude}
\end{figure}

There are two indications for the importance of the narrow passage area. 
The first one can be seen in figure \ref{fig:amplitude} which shows that 
$\langle v_y \rangle $ has (at all densities) a local maximum when the shaking amplitude $x_0$ is approximately half of the 
width of the passage. It is also interesting to note that only in the in the cooperative regime the 
average drift speed increases systematically with the amplitude $x_0$.   

\begin{figure}[t]
\begin{centering}
\includegraphics[width=8.8cm]{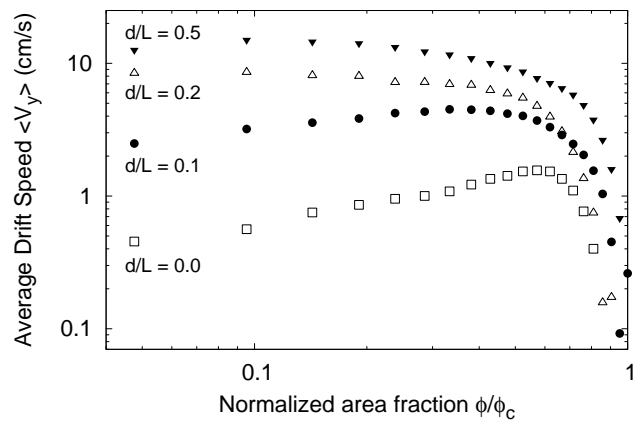}
\end{centering}
\caption{
Shifting the two sidewall by a distance $d/L$ as displayed in figure \ref{setup:shift}
increases the average drift velocity by an order of magnitude and removes the transition between
the dilute and the cooperative regime.
Numerical data measured with an oscillations amplitude of  7.5 mm and a driving frequency of 5 Hz.
}
\label{fig:shiftwall}
\end{figure}

The second sign that improved overcoming of the narrow passage is the mechanism behind 
the cooperative regime can be found when studying a channel where the two sidewalls are shifted 
by a length $d$ against each other as in figure  \ref{setup:shift}.  Figure \ref{fig:shiftwall}
clearly demonstrates that the distinction between dilute and cooperative regime vanishes 
when the ratio $d/L$ goes to 0.5 i.e.~the narrow passages are taken out of the system. 
The accompanying increase of the average drift speed by one order of magnitude is also significant from 
the perspective of potential applications.

\subsection{Jamming regime}
\begin{figure}[h]
\begin{centering}
\includegraphics[width=8.8cm]{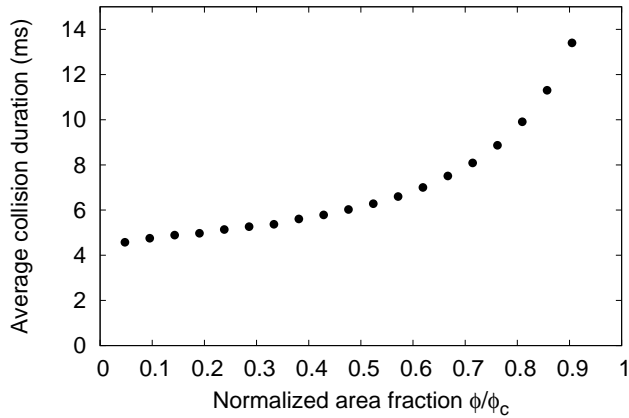}
\end{centering}
\caption{
The average collision duration doubles in the jamming regime, this hints towards the beginning formation 
of force-chains.
These numerical data are measured in the standard setup (figure \ref{setup:simu})
with an oscillations amplitude of  7.5 mm and a driving frequency of 5 Hz.
}
\label{fig:collisions}
\end{figure}

The strong decrease in drift velocity for $\phi/\phi_c > 0.6$ can be explained back to two mechanisms. First the effective redirection
of the kinetic energy at the sidewalls stops. As can be seen in figure \ref{fig:dist-alpha} 
at $\phi/\phi_c$ = 0.857 only 49 \% of the post collisional velocities have $\gamma$ values larger than 90 degree.

Secondly, figure \ref{fig:collisions} shows that  the average collision duration doubles in the jamming regime. 
For Hertzian single particles the collision duration is proportional to $v_n^{-0.2}$ \cite{schwager:98}.
As the kinetic energy of the particles decreases only by about 25\% in the jamming regime, this dependence
can not explain the observed increase in the average collision duration. We hypothesize therefore that 
the emergence of short-lived but system-spanning force chains is responsible for this increase.

\section{Conclusions}
We have demonstrated experimentally and numerically a new method of granular transport in a horizontally vibrated 
channel with sawtooth shaped boundaries. The average drift speed depends on the filling fraction, increasing 
for intermediate densities and decreasing again when approaching jamming. 
The underlying driving mechanism was found to be the redirection of the injected kinetic energy in the particle 
sidewall collisions. The reason for the drift speed increase at intermediate densities seems to be the improved 
flux through the narrow passages between the tips of the saw-teeth; removing them by introducing a phase shift between
the two boundaries resulted in a more than tenfold increase in drift velocity. At high densities the average 
contact duration increases which indicates that the energy injection through the shaking bottom plate becomes
significantly less efficient. However, more work will be needed for establishing a microscopic theory of 
this transport
mechanism.

{\bf ACKNOWLEDGMENTS}
We thank Y. Rahmani for his help in preparing the experimental setup and H. Fazli for helpful discussions.
Support from the IASBS Research Council (Grant No. G2012IASBS103 and G2012IASBS106 ) is acknowledged.

\bibliographystyle{apsrev}
\bibliography{ratchet_zanjan}

\end{document}